\PassOptionsToPackage{dvipsnames, table}{xcolor} 
\documentclass[sigconf]{acmart}

\usepackage{booktabs} % For formal tables
\usepackage[dvipsnames]{xcolor}
\usepackage{subcaption} % subfigure
\usepackage{multirow}
\usepackage{threeparttable} % footnote in table
\usepackage{makecell} % newline in cell
\usepackage{mathtools} % curly bracket in table
\usepackage{setspace} % setstretch
\usepackage{enumitem} % set indent gap for itemize and enumerate
\setlist[itemize,1]{leftmargin=12pt}
\setlist[enumerate,1]{leftmargin=16pt}
\usepackage[linesnumbered,algoruled,boxed,lined]{algorithm2e}
\renewcommand{\SetProgSty}[1]{\renewcommand{\ProgSty}[1]{\textnormal{\csname#1\endcsname{##1}}\unskip}}%
\SetProgSty{textsc}
\SetCommentSty{mycommentfont} % comment style

\ifdefined\sectionname
\renewcommand{\sectionname}{Section}
\else
\newcommand{\sectionname}{Section}
\fi

\copyrightyear{2024}
\acmYear{2024}
\setcopyright{acmlicensed}\acmConference[RACS '24]{International Conference on Research in Adaptive and Convergent Systems}{November 5--8, 2024}{Pompei, Italy}
\acmBooktitle{International Conference on Research in Adaptive and Convergent Systems (RACS '24), November 5--8, 2024, Pompei, Italy}
\acmDOI{10.1145/3649601.3698722}
\acmISBN{979-8-4007-0606-6/24/11}

\begin{document}

\title[Towards Building Private LLMs: Exploring Multi-Node Expert Parallelism on\\Apple Silicon for Mixture-of-Experts Large Language Model]{Towards Building Private LLMs: Exploring\\Multi-Node Expert Parallelism on Apple Silicon for\\Mixture-of-Experts Large Language Model}
\titlenote{This paper is a revised version of our work that received the Best Paper Award at ACM RACS 2024. The original version has not yet been indexed in the online proceedings. Readers can refer to it via the DOI link provided at the bottom-left corner of this page.\\}

%% ----------------------------------------------------------------
%% Author
%% ----------------------------------------------------------------
\author{Mu-Chi Chen}
\authornote{\scriptsize Also\,with Mohamed\,bin\,Zayed\,University\,of\,Artificial\,Intelligence,\,Abu\,Dhabi,\,United\,Arab\,Emirates.}
\orcid{0009-0007-6013-4122}
\affiliation{%
  \department{Dept. of Computer Science and Information Engineering}
  \institution{National Taiwan University}
  \city{Taipei}
  \country{Taiwan}
}
\email{muchi674@gmail.com}

\author{Po-Hsuan Huang}
\authornotemark[2]
\authornote{\scriptsize Also with Dept. of Computer Science and Information Engineering, National Taiwan University.}
\orcid{0000-0002-7458-9634}
\affiliation{%
  \department{Dept. of Computer Science and Information Engineering}
  \institution{National Cheng Kung University}
  \city{Tainan}
  \country{Taiwan}
}
\email{aben20807@gmail.com}

\author{Xiangrui Ke}
\orcid{0000-0003-1614-5688}
\affiliation{%
  \department{Dept. of Computer Science}
  \institution{Mohamed bin Zayed University of Artificial Intelligence}
  \state{Abu Dhabi}
  \country{United Arab Emirates}
}
\email{xiangrui.ke@mbzuai.ac.ae}

\author{Chia-Heng Tu}
\orcid{0000-0001-8967-1385}
\affiliation{%
  \department{Dept. of Computer Science and Information Engineering}
  \institution{National Cheng Kung University}
  \city{Tainan}
  \country{Taiwan}
}
\email{chiaheng@ncku.edu.tw}

\author{Chun Jason Xue}
\orcid{0000-0002-6431-9868}
\affiliation{%
  \department{Dept. of Computer Science}
  \institution{Mohamed bin Zayed University of Artificial Intelligence}
  \state{Abu Dhabi}
  \country{United Arab Emirates}
}
\email{jason.xue@mbzuai.ac.ae}

\author{Shih-Hao Hung}
\authornote{\scriptsize Also with High Performance and Scientific Computing Center, National Taiwan University.}
\authornotemark[2]
\orcid{0000-0003-2043-2663}
\affiliation{%
  \department{Dept. of Computer Science and Information Engineering}
  \institution{National Taiwan University}
  \city{Taipei}
  \country{Taiwan}
}
\email{hungsh@csie.ntu.edu.tw}

\renewcommand{\shortauthors}{M.-C. Chen et al.}

%% ----------------------------------------------------------------
%% Abstract
%% ----------------------------------------------------------------

\begin{abstract}
Large Language Models (LLMs) have revolutionized Artificial Intelligence (AI) with significant advancements such as OpenAI's ChatGPT, Meta's Llama, and Databricks' DBRX. This paper addresses the cost and scalability challenges encountered when constructing private LLM systems for personal or small group services, as aimed by Apple Intelligence. A Mac Studio cluster with Apple's M2 Ultra chips is established as a cost-efficient solution to host and accelerate the pretrained DBRX model with the Mixture-of-Experts (MoE) architecture. Our performance analysis reveal that parallel execution of the model's experts across two to four machine nodes significantly reduces inference time. We find that computation time for the experts is comparable to the communication time for exchanging their outputs, emphasizing the importance of network latency over bandwidth. We also observe significant management overhead due to Apple software stack's memory management logic. Based on these findings, we develop optimization schemes to eliminate the memory management overhead. As a result, the Mac Studio cluster is 1.15 times more cost-efficient than the state-of-the-art AI supercomputer with NVIDIA H100 GPUs. In addition, we construct a performance model to estimate system performance under varying configurations, and the model provides valuable insights for designing private LLM systems.
\end{abstract}


{\small
}


\maketitle

%% ----------------------------------------------------------------
%% Introduction
%% ----------------------------------------------------------------
\section{Introduction}

Large Language Models have emerged as an important development in the field of Artificial Intelligence. The advent of OpenAI's ChatGPT~\cite{chatgpt, brown2020language} marked a new milestone for AI. An example application, AI assistants, demonstrates the effectiveness of LLMs.
Numerous major and emerging companies are investing in the development of larger, more sophisticated LLMs. 
For instance, Meta's Llama~\cite{llama1, llama2}, a series of open pretrained models, has broadened the application domains of LLMs. Databricks' DBRX~\cite{dbrx}, an LLM with a staggering 132 billion parameters, sets a new benchmark for open-source models by leveraging the Mixture-of-Experts architecture. These advancements highlight the growing importance and potential of LLMs across various applications.

As LLMs continue to evolve, the private sector has recognized the value of developing in-house capabilities to run and manage these models, encouraging a movement to build private systems for LLMs, e.g., Apple Intelligence services. 
Such an approach offers several benefits. First, private LLM systems enhance data privacy by keeping data control within the organization. 
Second, private systems provide the opportunity to customize for specific needs. Third, system owners can have control over computational resources, which enables efficient resource allocation and management. Lastly, the ability to continuously refine the models ensures that they remain up-to-date and effective. 

Unfortunately, building private systems comes with its own challenges, such as the costs of establishing and maintaining the system, data collection and model training for specific needs, ethical considerations, and ensuring content quality. Especially, building and maintaining a private system for LLMs can be costly, as it involves substantial investment in hardware, software, and human resources. While high-performance computing devices and high-speed interconnect networking are often the preferred choices for building system hardware, they come with high costs and require experienced engineers to harness the underlying computing power. Furthermore, as the size and complexity of LLMs increase, system owners face the challenge of scalability. They must ensure that their systems can grow with the complexity of LLMs. Scaling the systems to handle larger models and datasets can be difficult. % research problem: high cost and scalability issues.

In this work, we aim to tackle the high cost and scalability issues associated with constructing private systems for LLMs. One of the key factors contributing to these issues is the hardware infrastructure required to support such a complex workload. Traditional hardware solutions often fall short of providing required computation and memory resources in a cost-efficient way. This is where Apple's workstation with its proprietary chips, known as Apple Silicon, offers a promising alternative\footnote{An Apple Mac Studio is far cheaper than an NVIDIA DGX GH200 supercomputer, and it has a smaller form factor than a commodity PC.}. Notably, the M2 Ultra~\cite{apple_m2ultra} chip encapsulates up to 24 CPU cores and 76 GPU cores. The Mac Studio, a small-form-factor workstation equipped with the M2 Ultra chip and up to 192 GB of unified, shared memory, is able to host many LLMs, such as Llama 2 70B. A larger LLM is possible to run on a cluster of Mac Studios connected with a 10 Gb Ethernet network.

We establish a Mac Studio cluster to run the pretrained, unquantized DBRX model with 132 billion bfloat16 parameters.
A series of experiments have been conducted to characterize the MoE model's performance running on Apple's software/hardware stack. We use a model implementation that leverages the MLX~\cite{Hannun_mlx}/Metal software to accelerate computation on M2 Ultra's GPUs. 
Our performance analysis results in Section~\ref{sec:analysis} can be categorized into three aspects.
\begin{itemize}
  \item The \emph{parallel execution} of the experts of the DBRX model across machine nodes shortens the model inference time. 
  \item The measured \emph{computation time} for the experts is comparable to the \emph{communication time} for exchanging their outputs, in terms of several milliseconds per model layer. To reduce the communication time, the network latency is more important than its bandwidth.
  \item The Apple's software stack presents a \emph{memory management} logic to orchestrate the execution of GPU programs in the environment of the unified memory (by using the on-demand weight loading), and the management overhead is significant for the execution of the DBRX model. 
\end{itemize}

Based on our findings, we implement different optimization schemes in Section~\ref{sec:method} 
to shorten the delay caused by the on-demand memory management. In addition, based on our performance analysis, we construct a performance model in Section~\ref{sec:evaluation} to estimate a performance bound for the cluster system, where the model reports the estimated model inference time under different numbers of machine nodes and different network configurations. This performance model can help the owners (or designers) of the private systems for the DBRX model to make better design decisions. The contributions made in this work are summarized as follows. 

\begin{enumerate}\small
  \item We establish a computer cluster consisting of Mac Studios for running the DBRX model with the MoE architecture. Substantial performance acceleration is achieved by the parallelized execution of the unquantized DBRX model on the Mac Studio cluster. This is an encouraging result for constructing private LLM systems. 
  \item We characterize DBRX's inference performance on Apple's software stack, such as the MLX and Metal frameworks. Our analysis results show that these software frameworks can have a significant impact on the performance of model inference, because of the activities made by the internal memory management logic. Optimization strategies are proposed to avoid the overhead caused by the internal software logic. 
  \item We develop a performance model to estimate a performance bound of the Mac Studio cluster on the DBRX model, according to the characterized performance on such a system. This performance model is useful for informed decision-making during the design phase of a private LLM system. 
  \item We show that our developed LLM system is 1.15 times more cost-efficient than an AI supercomputer powered by NVIDIA H100 GPUs, in terms of \emph{throughput per USD}. Especially, our built system achieves a throughput of 6.1 tokens/sec, and this is achieved during token generation. We believe the delivered performance is acceptable for most general use cases.
  \item We are not aware of any other work that can 1) construct a cluster system based on Mac Studios for running the unquantized DBRX model with the MoE architecture, 2) analyze the performance characteristics of such a system, 3) develop the optimization methods to accelerate the parallel execution of the experts across Mac Studios, and 4) create a performance model for estimating the performance bound for the systems built with varying numbers of machine nodes and different networking configurations.
\end{enumerate}

%% ----------------------------------------------------------------
%% Background
%% ----------------------------------------------------------------
\section{Background}

For the background of this work, we use Section~\ref{sec:moe} to introduce the Mixture-of-Experts architecture. Section~\ref{sec:parallel} introduces different parallelization techniques to distribute the model across multiple devices and leverage their resources for better performance.

\subsection{Mixture-of-Experts}\label{sec:moe}

\begin{figure}[ht]
  \vspace{-10pt}
  \centering
  \begin{subfigure}[b]{0.326\linewidth}
      \centering
      \includegraphics[width=\linewidth]{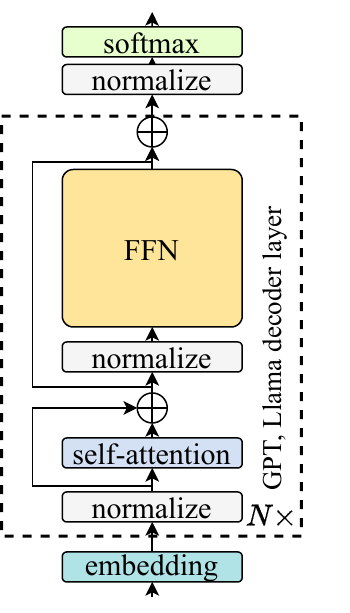}
      \captionsetup{skip=-1pt}
      \caption{Dense model.}\label{fig:gpt_llama}
  \end{subfigure}
  % \hspace{20pt}
  \begin{subfigure}[b]{0.423\linewidth}
      \centering
      \includegraphics[width=\linewidth]{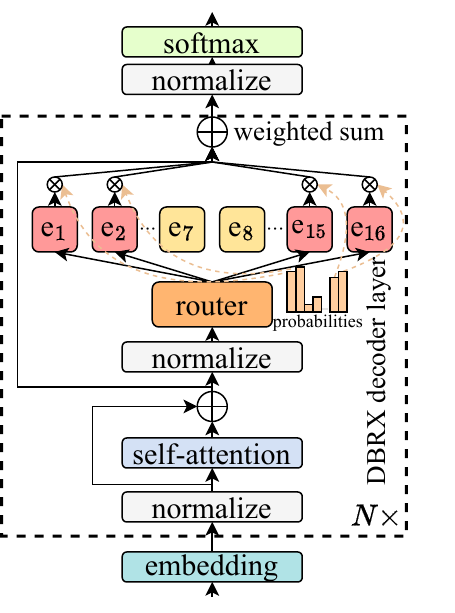}
      \captionsetup{skip=-1pt}
      \caption{MoE-based sparse model.}\label{fig:dbrx}
  \end{subfigure}
  \captionsetup{skip=2pt,font=small}
  \caption{Transformer-based decoder-only LLM architectures: dense model and MoE-based sparse model.}\label{fig:gpt_llama_vs_dbrx}
  % \vspace{-18pt}
\end{figure}

As shown in Figure~\ref{fig:gpt_llama_vs_dbrx}, prominent models like GPT~\cite{brown2020language} and Llama~\cite{llama1,llama2} (Figure~\ref{fig:gpt_llama}) are layers of transformer decoder blocks with two main components: the self-attention mechanism and a feedforward network (FFN). Instead of having just one FFN per layer, DBRX adopts the Mixture-of-Experts (MoE)~\cite{first_moe, lepikhin2020gshard, nlp_moe} architecture (Figure~\ref{fig:dbrx})  which uses an ensemble of multiple independent FFNs called \emph{experts}. DBRX selects 4 out of 16 experts in each layer. Although only part of the model parameters are activated for each token, all model parameters still need to be loaded into memory.

\subsection{Parallelization Techniques}\label{sec:parallel}

\emph{Tensor parallelism} partitions the tensor, or array, to different GPU processes. This technique relies largely on the fact that matrix multiplications can be broken down to a set of independent operations on independent data \cite{shoeybi2020megatronlm}. Here, input processing is potentially sped up because computation is parallelized. However, when met with tasks such as normalization, an all-reduce is required. Such tasks are frequent in LLMs, which makes tensor parallelism generally more suitable for single-node setups where GPU interconnect is very fast \cite{gpu_scale}. TensorRT-LLM~\cite{tensorrt_llm} is one of the leading frameworks that provides ample support for tensor parallelism, and DBRX~\cite{dbrx} also publishes an implementation using TensorRT-LLM.

\emph{Pipeline parallelism} range partitions a LLM's layers to different GPU processes. However, as each layer's input depends on the previous layer's output, a potential headache to this design is that only one GPU can work at a time. When the input batch size is greater than one, pipeline parallelism alleviates underutilization by breaking the batch into smaller micro-batches, which enables a GPU to continue working on the next micro-batch as it finishes the current one \cite{huang2019gpipe}. Note that these designs do not accelerate input processing \cite{miao2023efficient}. PyTorch currently supports pipeline parallelism with the PiPPy subpackage \cite{pippy2022}.

Specifically for the MoE architecture, \emph{expert parallelism} partitions experts to different GPU processes. This could improve performance because each expert is able to work on the input independently. In each layer, after the router selected experts are finished, an all-reduce is required for weighted sum, which means GPU interconnect speed is crucial to performance. Our distributed inference system uses expert parallelism because it requires less communication than tensor parallelism and still offers more performance benefits than pipeline parallelism with small batch sizes.

%% ----------------------------------------------------------------
%% Analysis
%% ----------------------------------------------------------------
\section{Analysis} \label{sec:analysis}
We established two parallel computing systems to assess the performance characteristics of the MoE model inference. The systems are quickly prototyped based on MLX's DBRX implementation\footnote{Code repository to MLX's DBRX reference implementation: \url{https://github.com/ml-explore/mlx-examples/blob/7d7e23606199b1224b628abbb5a0ba0cabd26398/llms/mlx_lm/models/dbrx.py}.}. The discovered performance results are reported in the following subsections. Further, our findings from the experimental analysis can be used to guide the optimizations to accelerate the model inference on parallel systems. 

\subsection{Parallel Model Inference with Four Nodes}
\begin{figure}[tbh!]
  \vspace{-10pt}
  \center
  \includegraphics[width=.8\linewidth]{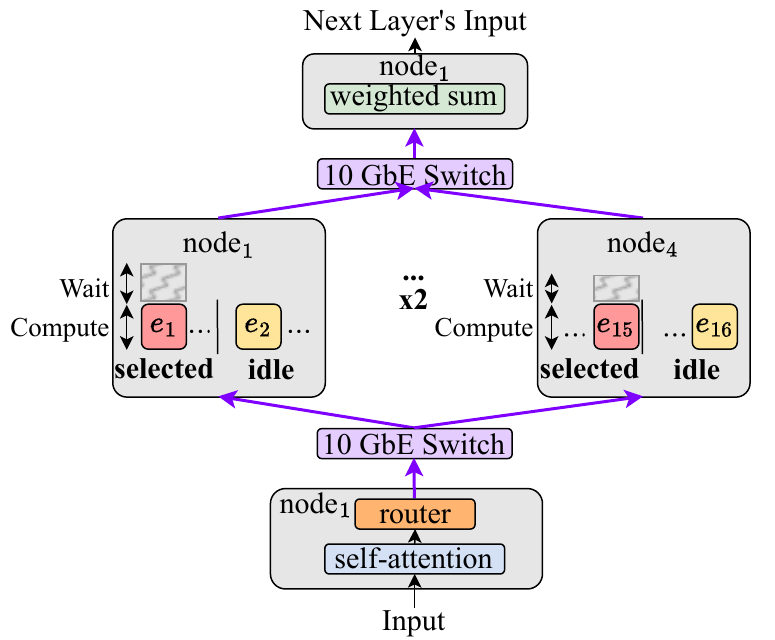}
  \captionsetup{skip=2pt,font=small}
  \caption{A parallel computing system organization with four nodes for a Mixture-of-Experts model of 16 experts. The figure illustrates the fork-join execution style for a single decoder layer of the model, and it is executed repeatedly.}
  \label{fig:general-arch}
  % \vspace{-10pt}
\end{figure}

A parallel computing system with four machine nodes is established to evaluate the performance of the DBRX model (a Mixture-of-Experts model with 16 experts), where the details of the machines and the model are provided in \sectionname~\ref{sec:evaluation}. As illustrated in \figurename~\ref{fig:general-arch}, the 16 experts are evenly distributed across the four machine nodes, with four experts on each node. When performing the model inference, the input data is fed to $node_1$ at the bottom of \figurename~\ref{fig:general-arch}, the four machine nodes take the output of the \emph{router} as their inputs for further processing (the odd experts are selected in this example), and the processing outputs of the experts are sent back to $node_1$ for further data aggregation. The aggregated outcomes at $node_1$ are served as the input for the next model layer, where the next iteration of the fork-join execution in \figurename~\ref{fig:general-arch} is performed. 

When performing model inference on each layer, each machine node is responsible for running the selected expert based on the output data provided by the router. It is important to note that the time spent on each node is dominated by the compute time for the selected expert. In addition, there is a synchronization point at $node_1$ to wait for the completion of all machine nodes; note that in \figurename~\ref{fig:general-arch}, the synchronization overhead is rendered as the wait time on each machine node for convenience, and the wait time includes data transfers and waiting for all machine nodes. The observations of the model inference on the four-node system are twofold. 
\begin{itemize}
  \item Partitioning experts to different nodes as shown in Figure~\ref{fig:general-arch} effectively reduces inference time by enabling experts to work in parallel. This realization of \emph{expert parallelism} is applicable to any number of nodes.
  \item The compute time is almost identical to the wait time for each layer of inference since the data to be computed and transferred is small (about 24,576 bytes), given that two selected experts are running per node. Based on this observation, we decide to run the inference with two machine nodes to lower the wait time (by reducing the number of communications per layer), leading to a minimized total execution time. In the following subsection, a set of experiments is conducted to discover the performance bottleneck during the model inference (i.e., to reduce the compute time).
\end{itemize}

\subsection{Parallel Model Inference with Two Nodes}\label{sec:twonodes}
\begin{figure}[tbhp]
  \vspace{-12pt}
  \center
  \includegraphics[width=.8\linewidth]{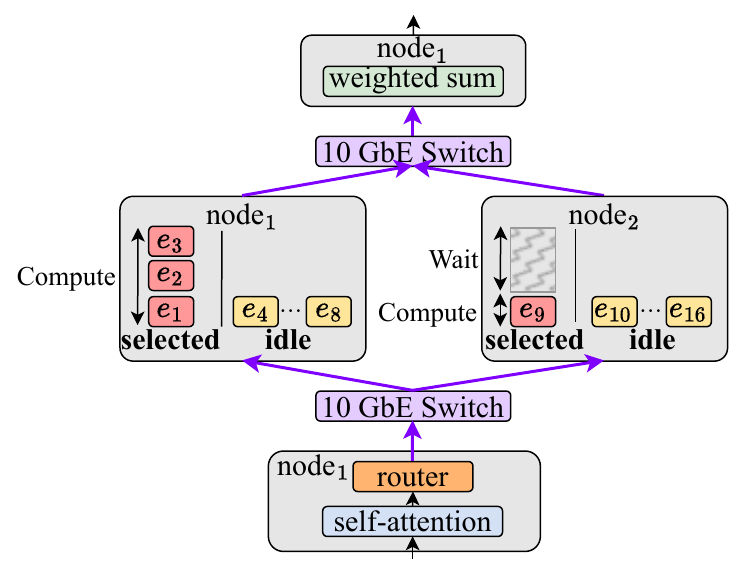}
  \captionsetup{skip=2pt,font=small}
  \caption{A two-node parallel system for the inference of the MoE model with 16 experts.}
  \label{fig:naive}
  \vspace{-8pt}
\end{figure}

A two-node parallel system is used to run the unquantized DBRX model, where the total memory of the two Mac Studio nodes (detailed in \sectionname~\ref{sec:evaluation}) is 384 GB. Each expert has roughly 7.9 billion parameters, and 16 experts account for 96\% of total weights for the entire model. \figurename~\ref{fig:naive} illustrates the 16 experts are distributed equally across the two nodes. As the DBRX model is built upon the MLX framework to run on Apple Silicon, Apple's performance profiler, Instruments, is adopted to collect performance statistics during inference. 

We measure the performance of the model, in terms of the compute time and wait time during the layer-by-layer inference. We find that while the wait time is reduced due to the fewer communications performed, the average compute time per expert is higher than that observed on the four-node system. Based on the performance data collected by Instruments, we learn that the Metal Driver spends a considerable amount of time wiring down data used by GPU's computation in the main memory. 
Since both the CPU and GPU share the main memory in Apple Silicon's unified memory architecture, we conjecture that said behavior is a protection mechanism against excessive GPU memory usage affecting CPU performance. 
On the other hand, the GPU driver has to make sure the required data are available and cannot be paged out before computation starts. 
Following the profiler's terminology, this data management behavior is referred to as \emph{\textbf{driver processing}} in the following content.

\begin{figure}[htbp]
  \vspace{-20pt}
  \center
  \captionsetup{skip=2pt,font=small}
  \includegraphics[width=.95\linewidth]{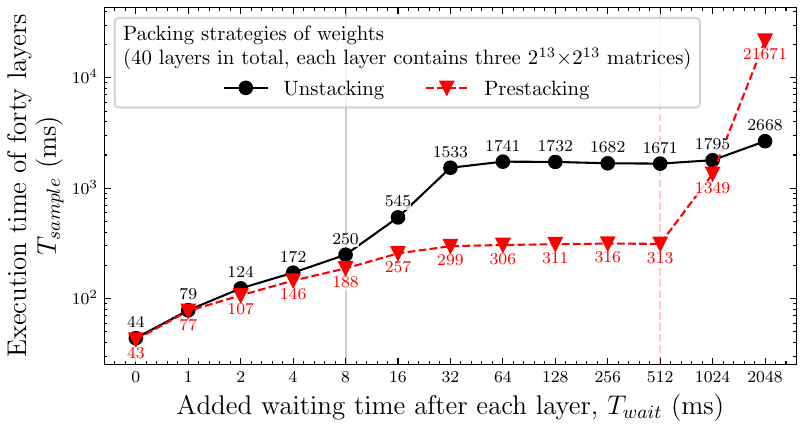}
  \caption{Analysis of the performance impact of the weight-packing strategies and the wait time on the execution time.}
  \label{fig:stack_sleep}
  \vspace{-8pt}
\end{figure}

A series of experiments are conducted to verify the existence of the driver processing behavior. A benchmark program is developed to mimic the behavior that occurred during the token generation stage of a single DBRX expert. The benchmark emulates the 40-layer model by 40 sets of three different matrix multiplications (Line 9 in Algorithm~\ref{algo:analysis}).
To test the driver's behavior, two weight-packing strategies are implemented as illustrated in Algorithm~\ref{algo:strategy}: 1) the unstacking approach which loads several 2D matrices from different files, and 2) the prestacking approach that loads a large 4D tensor into memory at once. These approaches use different methods to load the weights (into $B$, Line 4 in Algorithm~\ref{algo:analysis}).
The warmup process wires down all weight matrices beforehand, and the waiting time, $T_{wait}$, added by the \texttt{sleep} function (Line 22 in Algorithm~\ref{algo:analysis}) tests when the driver processing starts to reappear. 
\figurename~\ref{fig:stack_sleep} presents the benchmarking results, showing that compared to the unstacking strategy, prestacking maintains a more stable execution time until the added waiting time exceeds 512 ms. Our findings from these experiments are summarized as follows.

\begin{algorithm}[ht]
  \small
  \setstretch{0.56}
  \SetAlCapFnt{\small}
  \SetAlCapNameFnt{\small}
  \SetKwProg{Benchmark}{Benchmark}{ :}{end}
  \SetKwProg{Function}{Function}{ :}{end}
  \DontPrintSemicolon
  \caption{Two packing strategies of weights $B$.} \label{algo:strategy}
  $N_{\mathit{layers}}\gets 40$\tcp*[l]{Total number of layers}
  $N_{\mathit{mpl}}\gets 3$\ \ \ \ \tcp*[l]{Number of matrices per layer}
  $n\gets 8192$\;
  \BlankLine
  \Function{PackingStrategy($strategy$)}{
      \Switch{$strategy$}{
          \Case{Unstacking}{
              $B\gets\{\}$\;
              \For{$i\gets 1..N_{\mathit{layers}}$}{
                  \For{$j\gets 1..N_{\mathit{mpl}}$}{
                      $B_{i,j}$: \texttt{mx.array} ${\gets}_{load}$ $n{\times}n$ 2D matrix\;
                  }
              }
              \Return{$B$}\;
          }
          \BlankLine
          \Case{Prestacking}{
              $B$: \texttt{mx.array} ${\gets}_{load}$ $N_{\mathit{layers}}{\times}N_{\mathit{mpl}}{\times}n{\times}n$ 4D tensor\;
              \Return{$B$}\;
          }
      }
  }
\end{algorithm}

\begin{figure}[ht]
  \centering
  \begin{subfigure}[b]{\linewidth}
    \centering
    \includegraphics[width=.8\linewidth]{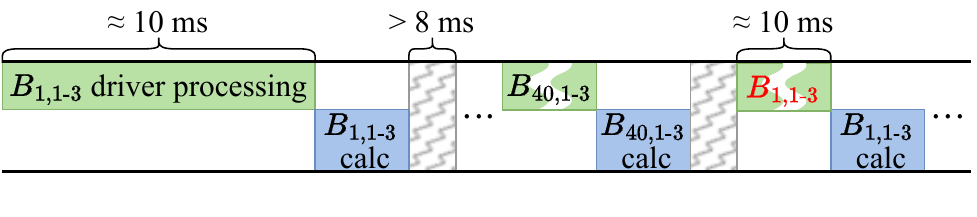}
    \captionsetup{skip=0pt}
    \caption{unstack driver processing}
    \label{fig:unstack_dp}
  \end{subfigure}

  \begin{subfigure}[b]{\linewidth}
    \centering
    \includegraphics[width=.8\linewidth]{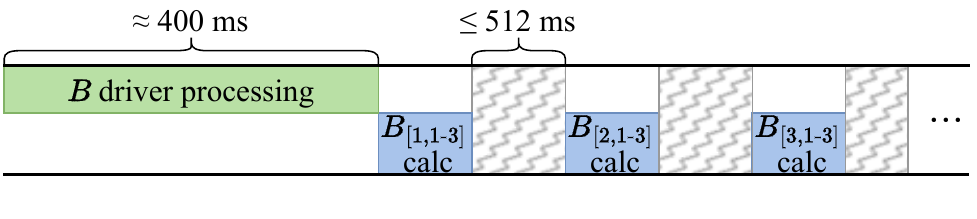}
    \captionsetup{skip=0pt}
    \caption{prestack no driver processing}
    \label{fig:prestack_no_dp}
  \end{subfigure}

  \begin{subfigure}[b]{\linewidth}
    \centering
    \includegraphics[width=.8\linewidth]{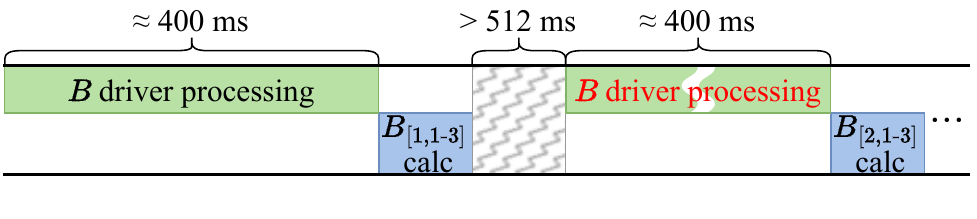}
    \captionsetup{skip=0pt}
    \caption{prestack driver processing}
    \label{fig:prestack_dp}
  \end{subfigure}
  \captionsetup{skip=2pt,font=small}
  \caption{Illustration of the runtime behaviors when running a layer of the benchmark with the unstacking and prestacking strategies. These timeline diagrams help understand the runtime activities of the experiments done in \figurename~\ref{fig:stack_sleep}.}
  \label{fig:dp_examples}
  \vspace{-5pt}
\end{figure}

\begin{itemize}
  \item \textbf{\emph{Finding 1.}} Considering the unstacking strategy, the average execution time per sample grows as we add more wait between layers; i.e., the added wait time ($T_{wait}$) is larger than 8 ms (which is indicated by the vertical black solid line in Figure~\ref{fig:stack_sleep}). In other words, there is a clear gap between the black line (the unstacking) and the red line (the prestacking) when $8 \leq T_{wait}  \leq 512$. This gap (time offset) is contributed by the driver preparing the required data for matrix multiplication. The timeline diagram for the operations performed during the execution of the benchmark is illustrated in \figurename~\ref{fig:unstack_dp}. Particularly, the previously wired weights $B_{1,1{\text -}3}$ are later wired again (the second $B_{1,1{\text -}3}$).

  \item \textbf{\emph{Finding 2.}} Considering the prestacking strategy, it requires a longer time (400 ms) initially for the driver to load the larger data, compared with the unstacking strategy. Nevertheless, the prestacking strategy results in a relatively stable execution time, especially when $8 \leq T_{wait}  \leq 512$, since the driver processing is not involved during the matrix multiplication performed on $B$, as depicted in \figurename~\ref{fig:prestack_no_dp}. It is important to note that the execution time grows exponentially as the added wait exceeds 512 ms, as shown on the right of \figurename~\ref{fig:stack_sleep}. Our profiling results indicate that the excessive time is from driver processing for the inference of each model layer, as illustrated in \figurename~\ref{fig:prestack_dp}.
\end{itemize}

\begin{algorithm}[ht]
  \small
  \setstretch{0.56}
  \SetAlCapFnt{\small}
  \SetAlCapNameFnt{\small}
  \SetKwProg{Benchmark}{Benchmark}{ :}{end}
  \SetKwProg{Function}{Strategy}{ :}{end}
  \SetKwFor{RepTimes}{repeat}{times do}{end}
  \DontPrintSemicolon
  \caption{Analysis of the effects of data packing strategies on matrix multiplication execution time.} \label{algo:analysis}
  $N_{\mathit{samples}}\gets 5$\;
  \BlankLine
  \Benchmark{(PackingStrategy)}{
      $A$: \texttt{mx.array} ${\gets}_{load}$ $1{\times}n$ vector\;
      $B\gets$ \textsc{PackingStrategy($\mathit{Unstacking}$ or $\mathit{Prestacking}$)}\;
      \For {$T_{\mathit{wait}} \gets 0, 2^0,2^1,2^2 ..., 2^{11}$}{
          \tcp{Warmup: wire down all needed memory}
          $A'\gets A$\;
          \For{$i\gets 1..N_{\mathit{layers}}$}{
              \For{$j\gets 1..N_{\mathit{mpl}}$}{
                  $A'\gets$ \texttt{mx.matmul(}$A', B_{i,j}$\texttt{)}\;
              }
              \texttt{mx.eval(}$A'$\texttt{)}\;
          }
          \tcp{Measure}
          $T_{\mathit{start}} \gets$ \texttt{now()}\;
          \RepTimes {$N_{\mathit{samples}}$}{
              $A'\gets A$\tcp*[l]{Reset $A'$}
              \For{$i\gets 1..N_{\mathit{layers}}$}{
                  \For{$j\gets 1..N_{\mathit{mpl}}$}{
                      $B'_{i, j}\gets B_{i, j}$ if $B$ is unstacked else $B_{[i, j]}$\;
                      $A'\gets$ \texttt{mx.matmul(}$A', B'_{i, j} $\texttt{)}\;
                  }
                  \texttt{mx.eval(}$A'$\texttt{)}\;
                  \texttt{sleep\_in\_milliseconds(}$T_{\mathit{wait}}$\texttt{)}\;
              }
          }
          $T_{\mathit{end}} \gets$ \texttt{now()}\;
          $T_{\mathit{sample}}\gets (T_{\mathit{end}}  - T_{\mathit{start}})/N_{\mathit{samples}} - (T_{\mathit{wait}}\times N_{\mathit{layers}})$
      }
  }
\end{algorithm}

It is important to note that while our benchmark is interfacing with the MLX framework, similar results are observed for the program built upon Metal. This indicates that the issue (our findings) may lie deeper in either the operating system or the GPU driver. However, determining the root cause is challenging due to the closed-source environment. We can only conclude that application developers should be aware of the above phenomenon when building applications using the same software stack, to avoid unexpected performance outcomes. 

%% ----------------------------------------------------------------
%% Methodology
%% ----------------------------------------------------------------
\section{Methodology}\label{sec:method}
Based on the parallel system design in \figurename~\ref{fig:naive} and our findings on driver processing in Section~\ref{sec:twonodes}, we develop three levels of optimizations. In Section~\ref{sec:pre}, we first prestack every layer's weights for each expert to help keep them all wired down. Second, to reduce the effect of experts unselected by the router being idle, we design one static and one dynamic load balancing strategy between nodes in Section~\ref{sec:loadingstrategies}. Third, we decentralize our system described in Section~\ref{sec:dec} to minimize internode communication time. Finally, to facilitate the design of private LLM systems, we build a performance model of our built system in Section~\ref{sec:bpm} to project a performance bound when the system is built with different configurations (i.e., number of machines and network interfaces).

\subsection{Expert-wise Weights Prestacking}\label{sec:pre}

With our naive implementation, each of the expert's weight matrices is loaded as a separate array. When not selected by the router, each layer's weights can sit idle for multiple tokens. Therefore, driver processing easily reappears. By prestacking every layer's weights, we greatly improve the data's chance of being referenced in computation. We do this through a one-time preprocessing script that stacks each expert's weights into a file and allows them to be loaded as a single array. In addition, we modified the expert to access its weights through indexing. While one could alternatively stack the separated arrays after loading them to achieve the same effect, such an approach creates a copy of the data when using MLX and thus doubles the memory usage. Meanwhile, we were also tempted to stack, on each node, all experts' weights together. This could not be done, however, because the aggregated size exceeds MLX's limit for a single array. 
Figure~\ref{fig:weights_warmup} illustrates two different weights warmup strategies that will be mentioned in the following subsections. In each layer, the red segments of each expert represent the active parameters in the prestacked weights.

\subsection{Multi-Node Compute Load Balancing} \label{sec:loadingstrategies}

\begin{figure}[t]
  \centering
  \begin{subfigure}[b]{0.31\linewidth}
      \centering
      \includegraphics[width=0.73\linewidth]{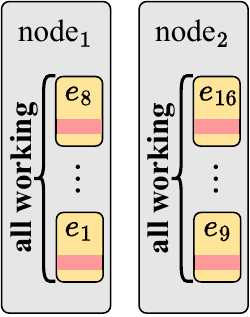}
      \captionsetup{width=\linewidth, skip=0pt}
      \caption{Busy Full Loading.}\label{fig:busy_full_loading}
  \end{subfigure}
  % \hspace{20pt}
  \begin{subfigure}[b]{0.6\linewidth}
      \centering
      \includegraphics[width=\linewidth]{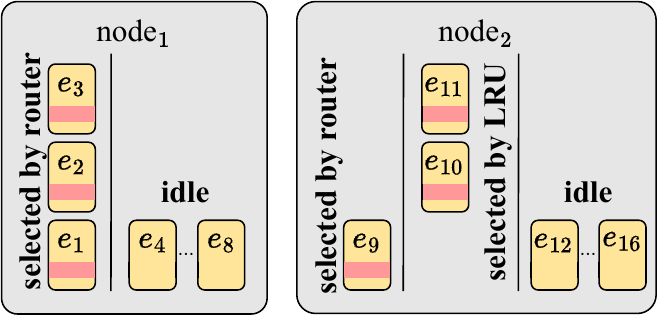}
      \captionsetup{width=\linewidth, skip=0pt}
      \caption{Router-Aided Dynamic Loading.}\label{fig:router_aided_dynamic_loading}
  \end{subfigure}
  \captionsetup{skip=2pt,font=small}
  \caption{Different Weights Warmup Strategies.}\label{fig:weights_warmup}
  \vspace{-8pt}
\end{figure}

\paragraph{Busy Full Loading}

With prestacking alone, we get trapped in a similar scenario as Figure~\ref{fig:prestack_dp}. Since only four of them are selected by the router in each layer, the 16 experts pay for their expensive first driver processing at separate times. In turn, this creates enough idleness to induce repeated payment, causing the system to enter an inescapable driver processing loop. Under the busy full loading strategy shown in Figure~\ref{fig:busy_full_loading}, we can avoid this by telling every expert to work every layer. For those that are not selected by the router, we zero out their response during the weighted sum. Albeit simple, this static load balancing strategy creates a lot of unused calculations as only 4 of the 16 computations spent are necessary.

\paragraph{Router-Aided Dynamic Loading}

To reduce wastage but still ensure that every expert is touched frequently enough, we propose the router-aided dynamic loading strategy. In each layer, the maximum number of experts selected by the router is calculated and sent to every node. In cases where the number of experts selected by the router on a node is less than the max, the spare computation quota goes to the \emph{least recently used (LRU)} experts. As shown in Figure~\ref{fig:router_aided_dynamic_loading}, red segments indicate the layer$_i$ of the prestacked expert weights. In this example, node$_2$ selects only one expert, which is fewer than the number selected by node$_1$. Therefore, two additional experts are chosen from the LRU to balance the load. According to our observation, this reduces our node's average load per layer from 8 in busy full loading to 2 experts. At the same time, our LRU mechanism ensures that each expert performs calculations in time before Metal Driver unwires their weights due to inactivity. Nonetheless, the mechanism above only works when all weights are already wired down. To satisfy this requirement, we pay all driver processing costs one-time at system startup and avoid repeated payment between inference requests by enforcing a standby calculation that sums up every expert's weights in GPU.

\begin{figure}[htbp]
  \vspace{-10pt}
  \center
  \includegraphics[width=.8\linewidth]{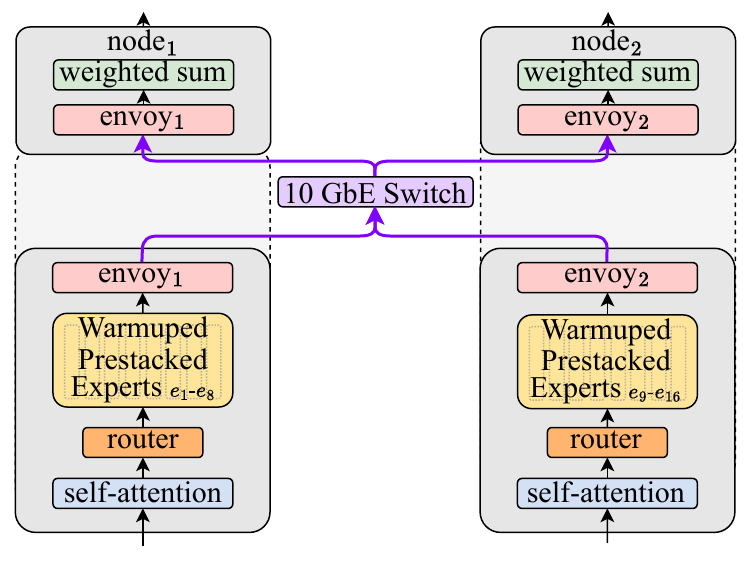}
  \captionsetup{skip=2pt,font=small}
  \caption{Architecture of one decoder layer with decentralized self-attention and router components.}
  \label{fig:decentralized}
  \vspace{-15pt}
\end{figure}

\subsection{Decentralized Self-Attention and Router}\label{sec:dec}

With the warmup strategies above in place, inter-node communication emerges as a performance bottleneck. Under Figure~\ref{fig:naive}'s architecture, each layer warrants 2 communications. With the inspiration from GShard~\cite{lepikhin2020gshard}, we are able to halve this by replicating the self-attention, router, and weighted sum components on every node. As demonstrated in Figure~\ref{fig:decentralized}, this eliminates the reliance on a central node for feeding experts' inputs and only keeps an unavoidable all-reduce on the experts' outputs. To handle the intermediate data sent and received during the all-reduce phase, we introduce an isolated process, \emph{envoy}, in each node. This process contains an asynchronous gRPC~\cite{grpc} server that acts as a dispatcher, minimizing disturbances to GPU computing. Note that the design introduces a slight increase in memory usage and repeated calculations, but node$_2$ in Figure~\ref{fig:naive} is idle anyway when node$_1$ is performing non-expert tasks. For our experiment setup, we consider this an acceptable tradeoff for the performance benefits.

\subsection{Performance Modeling}\label{sec:bpm}

From our experience deploying DBRX on parallel systems of different scales, we formulate our findings in Equation~(\ref{eq:pm}) as a performance model.
In our expert parallel systems that adopt the ${P{\text -}L_R{\text -}D}$ method, each node has equal load. Thus, in this setting, DBRX's lower bound inference time for each token during the token generation phase can be derived from one node's computation (\ref{eq:comp}) and communication (\ref{eq:comm}) time. The computation time is the maximum of how long the GPU spends loading computation data from memory to cache versus how long it spends calculating as these two can overlap. In most cases, the maximum is the load time. We estimate the size of the data that GPU needs to load to cache and the amount of floating number operations (FLOPs) happening in the calculation based on the mean number of experts working on a node during each layer. Data has to be reloaded to GPU cache every layer because each layer's weights (1.2 GB in a two-node system) far exceed our machine, M2 Ultra's, L3 cache size (96 MB). Next, the communication time is how long the layer-wise all-reduce operations took, which is the transport software (e.g., TCP/IP) processing latency plus data travel time. 
The variables (and their values) used in the equation are listed in Table~\ref{tab:vars}. The estimated execution time for different system configurations are provided in Section~\ref{sec:ceanalysis}.

\begin{subequations}
  \vspace{-8pt}
  \label{eq:pm}
  \small
  \begin{align}
    & \mathit{Est.\ Exec.\ Time}_{\mathit{\ lower\ bound}}\ = Max\Big(\nonumber\\
    \begin{split}
    &\underbrace{\frac{\mathit{\#Params_{SA}}+\mathit{\#Params_{/expert}}\times\mathit{\mathbb{E}[\mathit{\#exec.\ experts_{/node/layer}}]}}{\mathit{memory\ bandwidth}{/node}}}_{\mathit{GPU\ Load}}\\
    &, \underbrace{\frac{\mathit{FLOPs_{SA}}+\mathit{FLOPs_{/expert}}\times\mathit{\mathbb{E}[\mathit{\#exec.\ experts_{/node/layer}}]}}{\mathit{GPU\ BF16\ FLOPS{/node}}}}_{\mathit{GPU\ Compute}}\Big)
    \end{split}\label{eq:comp}\\ 
    &+ \Big(\underbrace{\mathit{comm.\ latency}\times\mathit{\#Layer}}_{\mathit{Comm.\ Latency}}+\underbrace{\frac{\mathit{comm.\ data}}{\mathit{comm.\ bandwidth}}}_{\mathit{Data\ Transfer\ Time}}\Big)\label{eq:comm}
  \end{align}
  \vspace{-10pt}
\end{subequations}

%% ----------------------------------------------------------------
%% Evaluation
%% ----------------------------------------------------------------
\section{Evaluation}
\label{sec:evaluation}

Section~\ref{sec:setup} outlines the experimental setup for building the Mac Studio cluster. Section~\ref{sec:perf} presents a performance breakdown of the proposed methods on the cluster system. Section~\ref{sec:scalability} analyzes the scalability of our best method on the system. 
Section~\ref{sec:ceanalysis} compares our solution with Databricks' in terms of cost efficiency. Section~\ref{sec:pm} reports the estimated performance of the cluster with different number of nodes and high-speed network interfaces. 

\subsection{Experimental Setup}\label{sec:setup}

Table~\ref{tab:setup} presents the hardware and software specifications of a single Mac Studio machine. The cluster systems are established using two to four Mac Studios, as illustrated in Section~\ref{sec:analysis}. As for the LLM model, the target MoE model for inference is the pretrained DBRX Instruct model with 132B parameters, with its architecture described in Section~\ref{sec:moe}.

\begin{table}[h]
  \vspace{-4pt}
  \begin{threeparttable}[b]
  \footnotesize
  \centering
  \captionsetup{skip=2pt,font=small}
  \caption{The performance model variables.}\label{tab:vars}
  \begin{tabular}{ll}
    \toprule
    \textbf{DBRX variable} & \textbf{Value}\\
    \;$\mathit{\#Layers}$ \& $\mathit{precision}$ & \phantom{0,0}40 \& \phantom{0,00}2 ($\because$ BF16)\\
    \;$D_\mathit{embed}$ \& $D_\mathit{qkv\_hidden}$ \& $D_\mathit{ffn}$ & 6,144 \& 8,192 \& 10,752\\
    \;$\mathit{comm.\ data}$ & \phantom{0}$2\times10^6$ bytes\tnote{(a)}\\
    \;\textbf{Self-Attention (SA)}: $\mathit{\#Params_{SA}}$ \& $\mathit{\#FLOPs_{SA}}$ & \phantom{0}$7\times 10^9$ bytes\tnote{(b)}\quad \& $14\times 10^9$\tnote{(c)}\\
    \;\textbf{MoE}: $\mathit{\#Params_{/expert}}$ \& $\mathit{\#FLOPs_{/expert}}$ & $16\times10^{9}$ bytes\tnote{(d)}\quad \& $16\times10^9$\tnote{(e)}\\
    \midrule
    \textbf{Hardware variable} & \textbf{Value}\\
    \;$\mathit{GPU\ BF16\ FLOPS{/node}}$ & \phantom{0}$54\times10^{12}$\\
    \;$\mathit{memory\ bandwidth}{/node}$ & $800\times10^9$ bytes/sec\\
    \;$\mathit{comm.\ bandwidth}$ & $1.25\times10^9$ bytes/sec\\
    \midrule
    \textbf{Independent variable} & \textbf{Value}\\
    \;$\mathit{\#Nodes}$ & $2, 3, 4$\\
    \midrule
    \textbf{Measured variable} & \textbf{Value}\\
    \;$\mathit{comm.\ latency}$ & $1 \times 10^{-3}$ sec\\
    \;\multirow{1}{*}{$\mathit{\#Nodes}$: $\mathbb{E}[\mathit{\#exec.\ experts_{/node/layer}}]$} & 2: 2.65 \& 3: 2.32 \& 4: 1.57\\
    \bottomrule
  \end{tabular}
  \begin{tablenotes}
    \footnotesize
    \item [(a)] $D_\mathit{embed}\times 4\times\mathit{\#Layers}\times\mathit{precision}\approx 2\times10^6$.
    \item [(b)] $(D_\mathit{qkv\_hidden}\times D_\mathit{embed}+D_\mathit{embed}^2)\times\mathit{\#Layers}\times\mathit{precision}\approx 7\times 10^9$.
    \item [(c)] $2\times \mathit{\#Params_{SA}}\approx 14\times 10^9$.
    \item [(d)] $D_\mathit{embed}\times D_\mathit{ffn}\times 3\times\mathit{\#Layers}\times \mathit{precision}\approx 16\times10^{9}$.
    \item [(e)] $2\times D_\mathit{embed}\times D_\mathit{ffn} \times \underbrace{3}_{\mathclap{v1,w1,w2}} \times \mathit{\#Layers}\approx 16\times10^9$.
  \end{tablenotes}
  \end{threeparttable}
  % \vspace{-24pt}
\end{table}

\begin{table}[h]
  \centering
  \small
  \captionsetup{skip=2pt,font=small}
  \caption{Hardware and software for a single Mac Studio node.}\label{tab:setup}
  \begin{tabular}{ll}
    \toprule
      \textbf{Parameter} & \textbf{Specification}\\
    \midrule
        Chipset & M2 Ultra\\
        CPU & 24-core (16 Performance + 8 Efficiency)\\
        GPU & 76-core (9,728 ALUs)\\
        Memory & Unified LPDDR5 192 GB (800 GB/s bandwidth)\\
        NIC & 10 Gigabit Ethernet Port\\ 
        OS & macOS Sonoma 14.5\\
        \multicolumn{1}{l}{\multirow{2}{*}{Software}} & Python 3.11.8, MLX v0.11.1, grpcio v1.62.2, \\
        & protobuf v4.25.3, and torch v2.2.2\\
        MoE model & unquantized DBRX Instruct 132B\\
    \bottomrule
  \end{tabular}
  % \vspace{-20pt}
\end{table}

\subsection{Performance Improvement}\label{sec:perf}

Table~\ref{tab:eval} presents the performance statistics of optimizations introduced in Section~\ref{sec:method} under a two-node setup. The naive version (Section~\ref{sec:twonodes}) simply parallelizes the DBRX model to get it running on the system, and hence it represents a baseline to compare against. ${P{\text -}L_B}$ and ${P{\text -}L_R{\text -}D}$ are combinations of different optimizations. $P$ stands for expert-wise weights \underline{P}restacking (Section~\ref{sec:pre}); $L_B$ and $L_R$ for multi-node compute \underline{L}oad balancing by \underline{B}usy full loading and \underline{R}outer-aided dynamic loading (Section~\ref{sec:loadingstrategies}); $D$ for \underline{D}ecentralized self-attention and router (Section~\ref{sec:dec}). During token generation, the naive version gives a 1.2 tokens/sec throughput. In comparison, ${P{\text -}L_B}$ which suppresses driver processing with extra expert calculations yields a 1.7x speedup in MoE execution time and a 2.1 tokens/sec throughput. Further, by reducing wasted calculation and communication time, ${P{\text -}L_R{\text -}D}$ achieves a 5.2x MoE speedup and a 6.1 tokens/sec throughput\footnote{Prompt evaluation throughput: 2.8 for naive, 4.8 for ${P{\text -}L_B}$, and 10.9 for ${P{\text -}L_R{\text -}D}$.}. Note that these results are profiled under a single-user workload, and input prompt and output generation are limited to 128 tokens.

\begin{table}[h]
  % \vspace{-4pt}
  \small
  \centering
  \captionsetup{skip=2pt,font=small}
  \setstretch{0.9}
  \caption{Token \underline{gen}eration throughput (TP.) and its per token performance breakdown of the proposed optimizations compared to the naive version. Miscellaneous includes self-attention, router, and weighted sum.}\label{tab:eval}
  \begin{tabular}{lc|c|ccc}
    \toprule
    \multicolumn{1}{c}{\multirow{2}{*}{\textbf{Method}}} & \multirow{2}{*}{\textbf{\makecell{gen TP.\\ {\footnotesize(tokens/sec)}}}} & \multirow{2}{*}{\textbf{\makecell{Time\\ {\footnotesize(sec/token)}}}} & \multicolumn{3}{c}{\textbf{\makecell{Breakdown {\footnotesize(sec/token)}}}} \\
    & & & \multicolumn{1}{c}{{\small\textbf{MoE}}} & \multicolumn{1}{c}{{\small\textbf{Comm.}}}  & \multicolumn{1}{c}{{\small\textbf{Misc.}}}\\
    \midrule
        Naive             & 1.2 & 0.857 & 0.378 & 0.357 & 0.122\\
        ${P{\text -}L_B}$ & 2.1 & 0.485 & 0.240 & 0.168 & 0.077\\
        ${P{\text -}L_R{\text -}D}$ & 6.1 & 0.166 & 0.081 & 0.038 & 0.047\\
    \bottomrule
  \end{tabular}
  \vspace{-4pt}
\end{table}

\subsection{Scalability}\label{sec:scalability}

To analyze the scalability of our work, we test our best method, ${P{\text -}L_R{\text -}D}$, on systems ranging from two to four nodes. As in Section~\ref{sec:perf}, the nodes here have the same specifications listed in Table~\ref{tab:setup} and are all connected to a 10 GbE switch. Also, input prompt and output generation are limited to 128 tokens. Table~\ref{tab:s_eval} breaks down the performance of the different-sized systems during token generation. With three and four nodes, the throughput increases from 6.1 in a two-node setup to 6.5 and 7.0 tokens/sec\footnote{Prompt evaluation throughput: 10.9 for two-nodes, 11.5 for three, and 13.6 for four.}. These improvements are brought predominantly by the decrease in MoE time. When adding more nodes to the system, we use the extra memory to load experts overlappingly. This enables us to distribute the computation more evenly among nodes. Meanwhile, notice how communication accounts for an increasing percentage (23\%, 29\%, and 33\%) of the generation time as the system grows from two to four nodes. This means, that despite our solution's ability to effectively cut MoE time with more nodes, its scalability remains greatly hindered by communication time.

\begin{table}[h]
  \vspace{-8pt}
  \small
  \centering
  \captionsetup{skip=2pt,font=small}
  \setstretch{0.9}
  \caption{Scalability of ${P{\text -}L_R{\text -}D}$ during token generation.}\label{tab:s_eval}
  \begin{tabular}{cc|c|ccc}
    \toprule
    \multicolumn{1}{c}{\multirow{2}{*}{\textbf{\#Nodes}}}& \multirow{2}{*}{\textbf{\makecell{gen TP.\\ {\footnotesize(tokens/sec)}}}} & \multirow{2}{*}{\textbf{\makecell{Time\\ {\footnotesize(sec/token)}}}} & \multicolumn{3}{c}{\textbf{\makecell{Breakdown {\footnotesize(sec/token)}}}} \\
    & & &\multicolumn{1}{c}{{\small\textbf{MoE}}} & \multicolumn{1}{c}{{\small\textbf{Comm.}}}  & \multicolumn{1}{c}{{\small\textbf{Misc.}}}\\
    \midrule
        2 & 6.1 & 0.166 & 0.081 & 0.038 & 0.047\\
        3 & 6.5 & 0.153 & 0.068 & 0.044 & 0.041\\
        4 & 7.0 & 0.144 & 0.054 & 0.048 & 0.042\\
    \bottomrule
  \end{tabular}
  \vspace{-20pt}
\end{table}

\begin{table}[h]
  \centering
  \small
  % \vspace{-10pt}
  \captionsetup{skip=2pt,font=small}
  \caption{Cost efficiency comparison.}\label{tab:comp_cost_eff}
  \begin{tabular}{lclcrr}
    \toprule
      \textbf{Solution} & {\small\textbf{$\mathit{\#Nodes}$}} & \textbf{\#GPUs$_{\mathit{/Node}}$} & {\small\textbf{\makecell{Price$_{\mathit{/Node}}$\\ (USD)}}} & \textbf{TP.} & \textbf{TP./USD} \\
    \hline
        Databricks & 1 & 8 ${\times}$ H100 & 289,000 & 112.5 & 0.000389\\
        Ours & 2 & 1 ${\times}$ M2 Ultra & \phantom{00}6,599 & \phantom{00}5.9 & 0.000447\\
    \bottomrule
  \end{tabular}
  \vspace{-20pt}
\end{table}

\subsection{Cost Efficiency}\label{sec:ceanalysis}

To establish our work's comparability, we evaluate our two-node system that adopts the ${P{\text -}L_R{\text -}D}$ method against Databricks' state-of-the-art solution in terms of cost efficiency in Table~\ref{tab:comp_cost_eff}. According to information on Databricks' GitHub\footnote{\setstretch{0.7}databricks/dbrx - How inference efficiency is measured \#9: {\tiny\url{https://github.com/databricks/dbrx/issues/9\#issuecomment-2025688421}}.}, they ran their experiments on an 8x-H100-80G system with NVIDIA TensorRT-LLM. For a fair comparison, we use the list price for both systems\footnote{\setstretch{0.7}M2 Ultra node list price: {\tiny\url{https://www.apple.com/shop/buy-mac/mac-studio/24-core-cpu-60-core-gpu-32-core-neural-engine-64gb-memory-1tb}}. Inferred 8x-H100-80G system specifications and list price from Arc Compute: {\tiny\url{https://web.archive.org/web/20240525002112/https://www.arccompute.io/solutions/hardware/customize-your-configuration-borealis-h100-server}}.}, though bulk purchases can reduce costs for both. Our token generation throughput results are from the same workload as Databricks' - single-user with 2000 input and 256 output tokens. They are slightly lower than in Table~\ref{tab:s_eval} because longer inputs require more computation during self-attention. Despite having a 22x cheaper setup, we managed to surpass Databricks in throughput per USD. Generating 5.9 tokens per second, our solution tunes a much more affordable setup to offer acceptable performance for private use.

\subsection{Performance Modeling}\label{sec:pm}

\paragraph{Est. perf. with different number of machines}

\begin{figure}[htbp]
  % \vspace{-20pt}
  \center
  \includegraphics[width=.75\linewidth]{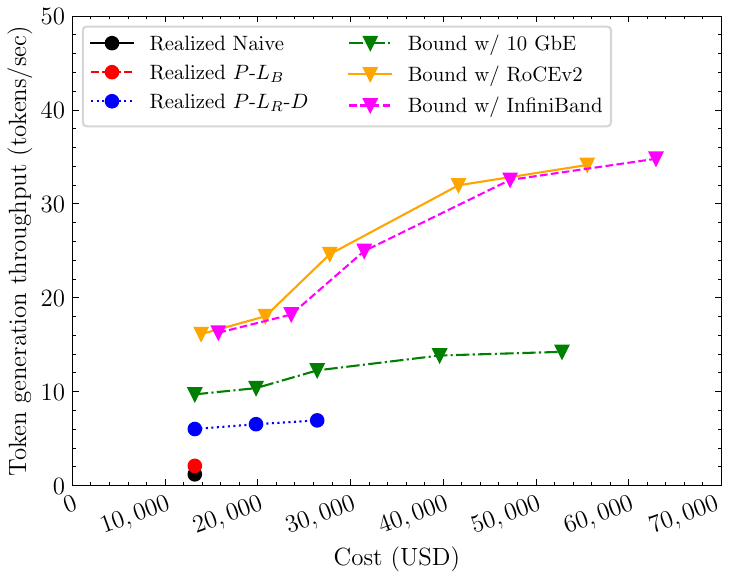}
  \captionsetup{skip=2pt,font=small}
  \caption{Realized results and theoretical performance bounds.}
  \label{fig:diff_comm}
  % \vspace{-14pt}
\end{figure}

Using Equation~\ref{eq:pm} and the data in Table~\ref{tab:vars}, Table~\ref{tab:predict} estimates the performance of our 10 GbE ethernet connected Mac Studio cluster from two to eight nodes. Columns on the left break down the time spent on GPU load/compute and communications while the performance bound for time and throughput is presented on the right.
Figure~\ref{fig:diff_comm} plots the estimated upper bound throughputs as green triangles whereas the blue dots exhibit the realized performance with two to four nodes from Table~\ref{tab:s_eval}. The estimated and realized data points are close to each other and uniform in trend, thus validating our performance model presented in Equation~\ref{eq:pm}. For reference, results from other versions of our system under a two node setup are shown in red and block dots. Visible growth of the red and the blue dots from the black confirm the effectiveness of our optimizations.

\begin{table}[ht]
  % \vspace{-6pt}
  \small
  \centering
  \captionsetup{skip=2pt,font=small}
  \setstretch{0.85}
  \caption{Estimated performance bounds for our Mac Studio cluster scaling from two to eight nodes.}\label{tab:predict}
  \begin{tabular}{ccccc|cc}
  \toprule
  \multicolumn{1}{c}{\multirow{2}{*}{\textbf{\#}}} & \multicolumn{2}{c}{\small\textbf{GPU Ops {\footnotesize (sec)}}}& \multicolumn{2}{c|}{\small\textbf{Comm. {\footnotesize (sec)}}}& \multicolumn{2}{c}{\multirow{1}{*}{\small\textbf{Bound w/ 10 GbE}}} \\
  &\multicolumn{1}{c}{\small\textbf{Load}}   & \multicolumn{1}{c}{\small\textbf{Comp.}} & \multicolumn{1}{c}{\small\textbf{Lat.}} & \multicolumn{1}{c|}{\small\textbf{Trans.}} & \multirow{1}{*}{\small\textbf{\makecell{Time {\scriptsize (sec)}}}}& \multirow{1}{*}{\small\textbf{\makecell{TP. {\scriptsize (tokens/sec)}}}}\\
  \midrule
  2 & 0.061 & \phantom{<}0.001  & 0.040 & 0.002 & 0.103 & \phantom{0}9.7 \\
  3 & 0.055 & \phantom{<}0.001  & 0.040 & 0.002 & 0.096 & 10.4 \\
  4 & 0.040 & \phantom{<}0.001  & 0.040 & 0.002 & 0.081 & 12.3 \\
  6 & 0.031 & <0.001 & 0.040 & 0.002 & 0.072 & 13.9 \\
  8 & 0.029 & <0.001 & 0.040 & 0.002 & 0.070 & 14.2 \\
\bottomrule
\end{tabular}
% \vspace{-16pt}
\end{table}

\paragraph{Est. perf. with different network interface cards}
The Mac Studio cluster is constrained by its communication network, as reported in Section~\ref{sec:scalability}. Equation~\ref{eq:pm} is used to project the performance when the Mac Studio cluster is upgraded with high-speed network interface cards that support RDMA through Infiniband or RoCEv2. These network interfaces can decrease the transport software processing latency that dominates communication time when using TCP/IP (Table~\ref{tab:predict}) by several orders of magnitude (to sub-600ns for Infiniband and 750ns for RoCEv2 as claimed by their manufacturers). Shown in Figure~\ref{fig:diff_comm}, the token generation throughput for the two-node cluster can be improved from 9.7 (the left-most green triangle) to 16.3 tokens/sec (the left-most yellow and pink triangles) with only a small increase in cost (5\% with RoCEv2, 20\% with Infiniband)\footnote{\setstretch{0.7}RoCEv2 NIC - latency: 750 ns; bandwidth: 25 Gbps; price: 339 USD/NIC {\tiny(\url{https://web.archive.org/web/20231203200642/https://www.fs.com/products/119649.html})}. Infiniband NIC - latency: 600 ns; bandwidth: 200 Gbps; price: 1,267 USD/NIC {\tiny(\url{https://web.archive.org/web/20240416112730/https://www.colfaxdirect.com/store/pc/viewPrd.asp?idproduct=3669&idcategory=6})}.}, meaning a significant boost in cost efficiency. The data points also reveal that the cluster could scale better from two to eight nodes\footnote{For each color of the triangles, the five data points represent the estimated performance for clusters of 2, 3, 4, 6, and 8 nodes.}.

%% ----------------------------------------------------------------
%% Conclusion
%% ----------------------------------------------------------------
\section{Conclusion}

This work achieves efficient inference with a state-of-the-art MoE LLM on Apple Silicon clusters. We propose optimizations to warm up and prestack DBRX experts' weights to minimize idle time and wasted calculations. A decentralized design further improved efficiency by reducing communications, offering acceptable performance in a far more affordable environment than that of Databricks'. Using the proposed performance model, we demonstrate the potential of various hardware configurations for establishing cost-efficient private LLM systems. In the future, we aim to conduct deeper analysis such as incorporating the roofline model to help refine our methods. As a starting point, this work focuses on optimizing support for a single user. We are developing strategies to handle multiple concurrent users and other more complex needs.

%% ----------------------------------------------------------------
%% Acknowledgments
%% ----------------------------------------------------------------
\begin{acks}
  This work was supported in part by National Science and Technology Council, Taiwan, under Grants 112-2221-E-002 -159 -MY3 and 111-2221-E-006 -116 -MY3.
\end{acks}

\begingroup
\setstretch{0.95}
\bibliographystyle{ACM-Reference-Format}
{\tiny\bibliography{paper}}
\endgroup

\end{document}